\documentclass[prb,twocolumn,superscriptaddress,showpacs]{revtex4}% Physical Review Letters

\usepackage{graphicx}% Include figure files
\usepackage{dcolumn}% Align table columns on decimal point
\usepackage{bm}% bold math
\usepackage{subfigure}

\begin{document}

\newcommand{\DMO}{DyMn$_{2}$O$_{5}$}
\newcommand{\RMO}{{\it R}Mn$_{2}$O$_{5}$}

%\preprint{RAE-1}

\title{An x-ray resonant diffraction study of multiferroic
DyMn$_{2}$O$_{5}$}

\author{R. A. Ewings}
\email{r.ewings1@physics.ox.ac.uk} \affiliation{ Department of
Physics, University of Oxford, Oxford, OX1 3PU, United Kingdom }
\author{A. T. Boothroyd} \affiliation{ Department of
Physics, University of Oxford, Oxford, OX1 3PU, United Kingdom }
\author{D. F. McMorrow}\affiliation{London Centre for Nanotechnology
and Department of Physics and Astronomy, University College
London, London, WC1E 6BT, United Kingdom}
\author{D. Mannix}\affiliation{XMaS CRG Beam line, European
Synchrotron Radiation Facility, F-38043 Grenoble, France }
\author{H. C. Walker}\affiliation{London Centre for Nanotechnology
and Department of Physics and Astronomy, University College
London, London, WC1E 6BT, United Kingdom}
\author{B. M. R. Wanklyn} \affiliation{ Department of
Physics, University of Oxford, Oxford, OX1 3PU, United Kingdom }

\date{\today}

\begin{abstract}
X-ray resonant scattering has been used to measure the magnetic
order of the Dy ions below 40\,K in multiferroic
DyMn$_{2}$O$_{5}$. The magnetic order has a complex behaviour.
There are several different ordering wavevectors, both
incommensurate and commensurate, as the temperature is varied. In
addition a non-magnetic signal at twice the wavevector of one of
the commensurate signals is observed, the maximum intensity of
which occurs at the same temperature as a local maximum in the
ferroelectric polarisation. Some of the results, which bear
resemblence to the behaviour of other members of the {\it
R}Mn$_{2}$O$_{5}$ family of multiferroic materials, may be
explained by a theory based on so-called acentric spin-density
waves.

\end{abstract}


\maketitle

\section{Introduction}

In recent years there has been an upsurge in research into materials
which display coupling between ferromagnetic and ferroelectric order
parameters. Such materials, which are members of the class known as
multiferroics, have been known to science for some time
\cite{Old_MF_review,Cr2O3Dzyaloshinskii,AstrovCr2O3,RadoCr2O3,Al'shinTi2O3,RadoGaFeO3,FischerA2M4O9,HolmesDyAlO3},
however recent discoveries of compounds in which the
magneto-electric coupling is strong have sparked a renaissance. Of
particular interest are the {\it R}MnO$_{3}$ and {\it
R}Mn$_{2}$O$_{5}$ ({\it R} = rare earth) materials. Indeed, the
current revival of the multiferroic field was stimulated by
measurements on TbMnO$_{3}$ \cite{Kimura nature,Goto PRL}, which
shows an unprecedentedly strong magneto-electric coupling.

A common feature of these manganese oxide based compounds is a
complex magnetic phase diagram, including magnetic order that is in
turn antiferromagnetic, incommensurate and/or commensurate in
wavevector \cite{Kimura nature,Goto PRL, Kenzelmann PRL}. Another
common feature is the possibility to control ferroelectric (FE)
polarisation by the application of a moderate magnetic field. This
control can take the form of enhancing, creating, or switching the
direction of the polarisation. Several theories, based on symmetry
considerations, have been proposed to explain such phenomena
\cite{Mostovoy,Betouras}.

The bulk properties of the {\it R}Mn$_{2}$O$_{5}$ ({\it R} = Tb,
Ho, Er, Dy, ...) compounds have been studied in some detail and
have many similar features \cite{Hur nature,Hur paper,Higashiyam
Ho/Er,Higashiyama,Inomata Gd Tb and Y}. In zero applied magnetic
field there exists a finite FE polarisation along the $b$-axis in
the approximate temperature range $20 \leq T \leq 35$\,K, which
increases rapidly at the lower end of the temperature range and
decreases more gradually at the upper end. Such behaviour is
present for all {\it R}Mn$_{2}$O$_{5}$ compounds, however the
magnitude of the polarisation is largest in DyMn$_{2}$O$_{5}$, and
appears to be significantly weaker if {\it R} is a non-magnetic
ion such as Y \cite{Inomata Gd Tb and Y}. It would seem,
therefore, that the presence of a magnetic rare-earth enhances the
FE polarisation.

The temperature dependence of the FE polarisation of
DyMn$_{2}$O$_{5}$ is somewhat complicated \cite{Higashiyama}.
Below about 7K the polarisation is negligibly small, however for
$7 \leq T \leq 12$K the polarisation is small but no longer
negligible. Between 12K and 15K there is a sharp rise in
polarisation, with a maximum just above 15K. The polarisation then
steadily falls with increasing temperature up to about 22K. Above
this temperature the polarisation rises again, reaching a local
maximum (which is slightly lower than the polarisation at 15K) at
about 25K. Further increase of temperature results in a steady
reduction of polarisation, until it becomes zero at about 39K.
Higashiyama {\it et al} have labelled three different
ferroelectric phases as follows: FE3 for $7 \leq T \leq 14$K, FE2
for $14 < T \leq 27$K, and FE1 for $27 < T \leq 39$K. The upper
temperature limit of the polarisation is independent of applied
magnetic field, as is the existence of a sharp spike in the
dielectric permittivity. If a magnetic field is applied along a
particular crystallographic axis (which axis is material
dependent, it seems) then ferroelectric effects occur at lower
temperatures, including increased polarisation, a switching of the
direction of polarisation and increases in the dielectric
permittivity.

As well as the ferroelectric phase, other common features among
{\it R}Mn$_{2}$O$_{5}$ compounds are the existence of a magnetic
transition from an ordered to a disordered state at $T_{\rm N}
\sim 40$\,K, with a transition into a ferroelectric state just
below $T_{\rm N}$. At low temperatures, typically $T<10$\,K, the
magnetic rare earth ions are observed to order
antiferromagnetically. The precise details of these changes are
different, depending on the choice of {\it R}. This may depend on
the magnetic easy axis of the material, which is the $a$-axis for
{\it R} = Tb, the $b$-axis for {\it R} = Dy and Ho, and the
$c$-axis for {\it R} = Er and Tm. All of these transitions are
accompanied by distinct anomalies in the specific heat.

DyMn$_{2}$O$_{5}$ crystallises in the orthorhombic space group
$Pbam$ with the lattice parameters $a=7.294$\,{\AA},
$b=8.555$\,{\AA} and $c=5.688$\,{\AA}. The structure is such that
interspersed between sheets of Dy$^{3+}$ ions there are, in order
along the $c$-axis, a Mn$^{4+}$O$_{6}$ octahedron, a
Mn$^{3+}$O$_{5}$ bipyramid, followed by another Mn$^{4+}$O$_{6}$
octahedron \cite{Blake}.

There have been several neutron scattering experiments conducted on
DyMn$_{2}$O$_{5}$ \cite{Blake,Wilkinson,Ratcliff}. These experiments
have had some success in elucidating the crystallographic and
magnetic structure of this material. It appears that there are three
distinct magnetic phases present at various temperatures below
$T_{\rm N}$. Below $T\approx 8$\,K the Dy ions are modulated
antiferromagnetically (AFM) along the $a$-axis, with wavevector
$\mathbf{q}_{AFM}^{Dy} = (0.5,0,0)$, with their moments pointing
along the $b$-axis. From base temperature up to $T_{\rm N}$ the
Mn$^{3+}$ and Mn$^{4+}$ ions undergo several magnetic transitions
between an incommensurate magnetic (ICM) phase and a commensurate
magnetic (CM) phase. The propagation vectors of these phases will be
denoted hereafter by $\mathbf{q}_{ICM}$ and $\mathbf{q}_{CM}$
respectively, where $\mathbf{q}_{ICM}=(0.5 \pm \delta,0,0.25 \pm
\epsilon)$ and $\mathbf{q}_{CM}=(0.5,0,0.25)$. The existence of
these phases has been shown in all the neutron scattering
measurements, but the precise details of ordering wavevectors and
temperatures are not consistent between studies.

Wilkinson {\it et al.} \cite{Wilkinson} measured the magnetic
structure at 4.2\,K and found, coexisting with the AFM order of
the Dy ions, an ICM structure with $\delta = 0$ but $\epsilon \neq
0$. The value of $\epsilon$ was found to vary slightly in the
range $0\leq \epsilon \leq 0.002$ for the three temperatures
measured (4.2\,K, 18\,K and 44\,K). This incommensurate order was
attributed to the ordering of both sets of Mn ions. More recently,
neutron powder diffraction measurements by Blake {\it et al.}
\cite{Blake} were able to resolve a magnetic structure with
$\delta = 0.01$ and $\epsilon = 0$ below 32\,K, which they did not
observe to change on cooling until $T\leq8$\,K, whereupon it
became much weaker and an AFM phase the same as that seen by
Wilkinson {\it et al.} was found.

Subsequently, single crystal neutron diffraction measurements were
made by Ratcliff {\it et al.} \cite{Ratcliff} over a temperature
range $2$\,K$\leq T \leq 45$\,K. Like the other measurements they
found an AFM phase below 8\,K, but at higher temperatures they
found a more complex behaviour. Two different ICM phases were
found for $8$\,K$ \leq T \leq 18$\,K characterised by the
wavevector $\mathbf{q}_{ICM}=(0.5,0,0.25+\epsilon_{1,2})$, where
$\epsilon_{1,2}$ are two different incommensurabilities which vary
in size between 0 and 0.02 and have opposite sign. For $18$\,K$
\leq T \leq 33$\,K one of the ICM peaks becomes CM whilst the
other's intensity gradually decreases with increasing temperature.
For $T
> 33$\,K there exists only the CM phase, and on warming this
disappears by $\sim 40$\,K.

Up to now there have been no x-ray resonant scattering (XRS) studies
of DyMn$_{2}$O$_{5}$, although there have been two XRS studies of
the related compound TbMn$_{2}$O$_{5}$ \cite{Okamoto TbMn2O5,Koo
TbMn2O5}, both of which have concerned themselves with the ordering
of the Mn sublattice by tuning the incident x-ray energy to one of
the Mn edges. For $10 \leq T \leq 41$\,K a pair of ICM peaks at
$(0.5\pm \delta,0,0.25+\epsilon)$ are observed, where $\delta$
decreases from about 0.012 at 10\,K to 0.003 at 25\,K, where it
remains constant until 32\,K above which it increases to about 0.012
at 41\,K. The $c$-axis incommensurability $\epsilon$ gradually
decreases from 0.06 at 10\,K to 0.03 at 41\,K. The intensity of the
ICM peaks decreases steadily with increasing temperature. Above
about 21\,K a CM peak at $(0.5,0,0.25)$ appears, the intensity of
which increases with increasing temperature up to about 30\,K, and
then decreases on further warming to 41\,K. It is also noted that
the (3,0,0) Bragg peak, which is forbidden in the $Pbam$ space group
that characterises the crystal structure of TbMn$_{2}$O$_{5}$, is
observed to scale exactly as the FE polarisation squared. The
$(3,0,0)$ Bragg peak would be allowed if the space group became
non-centrosymmetric, and lack of a centre of inversion is required
in all of the proposed explanations for the occurrence of FE
polarisation in magnetically ordered materials.

There were several motivations for this XRS study of
DyMn$_{2}$O$_{5}$. Firstly, XRS offers the possibility of probing
the magnetic order on the Dy and Mn sites separately by using the
resonant enhancement in the scattering when the x-ray energy is
tuned to an atomic absorption edge. Neutron diffraction, by
contrast, is sensitive only to the size of the overall magnetic
moment and not to the atomic species to which it is attached.
Secondly, XRS offers a high wavevector resolution, which makes it
possible to measure changes in the magnetic ordering wavevector
with very high accuracy. Thirdly, XRS is particularly useful for
the case of DyMn$_{2}$O$_{5}$ because the neutron absorption cross
section for $^{164}$Dy is relatively large and this isotope makes
up about 28\% of naturally occurring Dy. Hence, it is difficult to
obtain a good signal to background ratio in a neutron diffraction
measurement on DyMn$_{2}$O$_{5}$. The advantages of XRS have
enabled us in this study to clarify the existing data on
DyMn$_{2}$O$_{5}$, and to provide a more detailed picture of the
various ordering features of this important multiferroic material.

\section{Experimental details}

The single crystal sample was grown using the flux method
\cite{Wanklyn crystal growth}. Preliminary x-ray measurements using
a standard Laue camera showed that the lattice parameters and space
group of the sample used were consistent with those expected from
the literature.

Magnetometry measurements were performed using a Quantum Design
SQUID magnetometer, and exhibited the same behaviour as previously
reported \cite{Hur paper}. Specific heat measurements were performed
in zero applied magnetic field using a Quantum Design PPMS. After
alignment of the crystal with the x-ray Laue the sample was cut and
polished so that the $(0,0,1)$ direction was normal to the surface.
Most of the synchrotron x-ray measurements were made at wavevectors
of the form $\mathbf{Q}=(0,0,4) \pm \mathbf{q}$.

\begin{figure}[!h]
\includegraphics*[scale=0.65,angle=0]{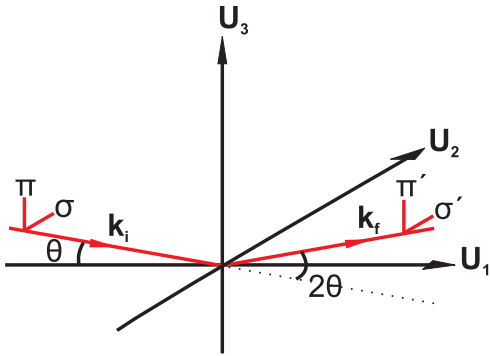}
\centering \caption{(Color online) The vertical x-ray scattering
geometry, defining U$_{1}$, U$_{2}$ and U$_{3}$, as well as the
polarisation directions and scattering angle
$2\theta$}\label{Scattering plane}
\end{figure}

X-ray resonant scattering (XRS) and non-resonant x-ray scattering
measurements were performed on the XMaS beamline at the European
Synchrotron Radiation Facility (ESRF), Grenoble, France. The
source of x-rays at XMaS is a bending magnet, so the light is
linearly polarised in the horizontal plane. A vertical scattering
geometry was used, shown in Figure \ref{Scattering plane}, so the
incoming x-rays were $\sigma$-polarised. For the XRS measurements
an incident energy of 7.795\,keV, corresponding to the Dy
L$_{3}$-edge (see below), was used. For non-resonant measurements
an incident energy of 11.17\,keV was used. For the XRS
measurements at the Dy L$_{3}$-edge a Au(222) analyser crystal
between the sample and detector was used in order to measure the
$\sigma'$ and $\pi'$ polarised components in the scattered beam.

The magnetic scattering amplitude in XRS, for each polarisation
state, is given by \cite{Hill and McMorrow}

\begin{eqnarray}
\mathbf{A}_{\mbox{res}}^{\mbox{Mag}} & = & \left(\begin{array}{cc}
                \sigma \rightarrow \sigma' & \sigma \rightarrow \pi' \\
                \pi \rightarrow \sigma' & \pi \rightarrow \pi'
                \end{array} \right) \\
                & = & F_{E1}^{(1)}\left( \begin{array}{cc}
                0 & z_{1}\mbox{cos}\theta + z_{3}\mbox{ sin}\theta \\
                z_{3}\mbox{ sin}\theta - z_{1}\mbox{ cos}\theta & -z_{2}\mbox{ sin}2\theta
                \end{array} \right) \nonumber
\end{eqnarray}

\noindent where z$_{1}$ is parallel to U$_{1}$ and so on (see
Figure \ref{Scattering plane}).

The Dy L$_{3}$-edge, which involves virtual electronic transitions
between the 2$p$ and 5$d$ states, was used for XRS measurements
because the resonant enhancement of the magnetic scattering at
this energy can be several orders of magnitude. This contrasts
with the resonant enhancement of magnetic scattering at the Mn
K-edge (1$s$ $\rightarrow$ 4$p$), which is only a factor of about
3. Furthermore, the spectrum of the bending magnet source at XMaS
is such that the flux of x-rays with energies near the Dy
L$_{3}$-edge is significantly higher than the flux of x-rays with
energies near the Mn K-edge. Thus one would expect it to be
possible to observe magnetic scattering arising from ordering on
both the Mn and Dy sublattices by measuring the signal from the Dy
sublattice alone, assuming that the magnetic polarisation of the
5$d$ states of the Dy ions is caused by a combination of the local
magnetic environment due to the Mn ions and the magnetisation of
the Dy 4$f$ electrons.

An energy of 11.17\,keV was chosen for the non-resonant
measurements for two reasons. First, at higher x-ray energies the
penetration depth of the x-rays is larger, thus increasing the
scattering volume and making the signal less surface sensitive.
This means that any observed signal is both more intense and
sharper in wavevector than it would be for lower energies. Second,
this energy is sufficiently high that although the flux of x-rays
from the bending magnet is still very high, the flux of higher
harmonic x-rays is vastly smaller. This effectively eliminates the
possibility of mistaking scattering at this energy with scattering
resulting from second or third harmonic x-rays.

\section{Results}

\begin{figure}[!h]
\includegraphics*[scale=0.4,angle=0]{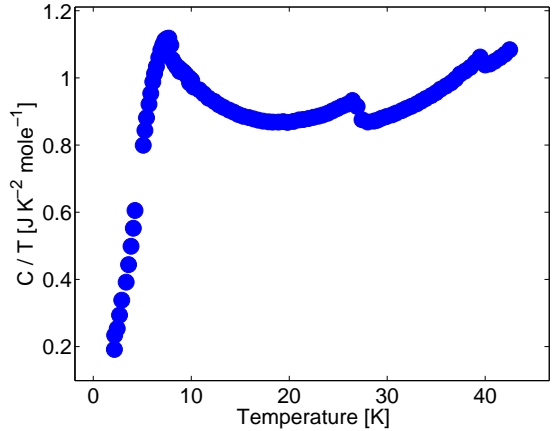}
\centering \caption{(Color online) Measurement of the specific heat
of the crystal of DyMn$_{2}$O$_{5}$ used in the x-ray measurements.
The temperatures at which there are anomalies agree with those
measured previously \cite{Hur paper,Higashiyama}. Note that the
lattice specific heat has not been subtracted from these
data.}\label{My meas HC}
\end{figure}

The specific heat of the sample used for this XRS study is shown
in Figure \ref{My meas HC}. There are clear anomalies at
$T=7.2$\,K, 27\,K and 39\,K, which is in agreement with previous
measurements of the specific heat \cite{Higashiyama}.

\begin{figure}[!h]
     \centering
     \subfigure{
          \includegraphics*[angle=270,scale=0.32]{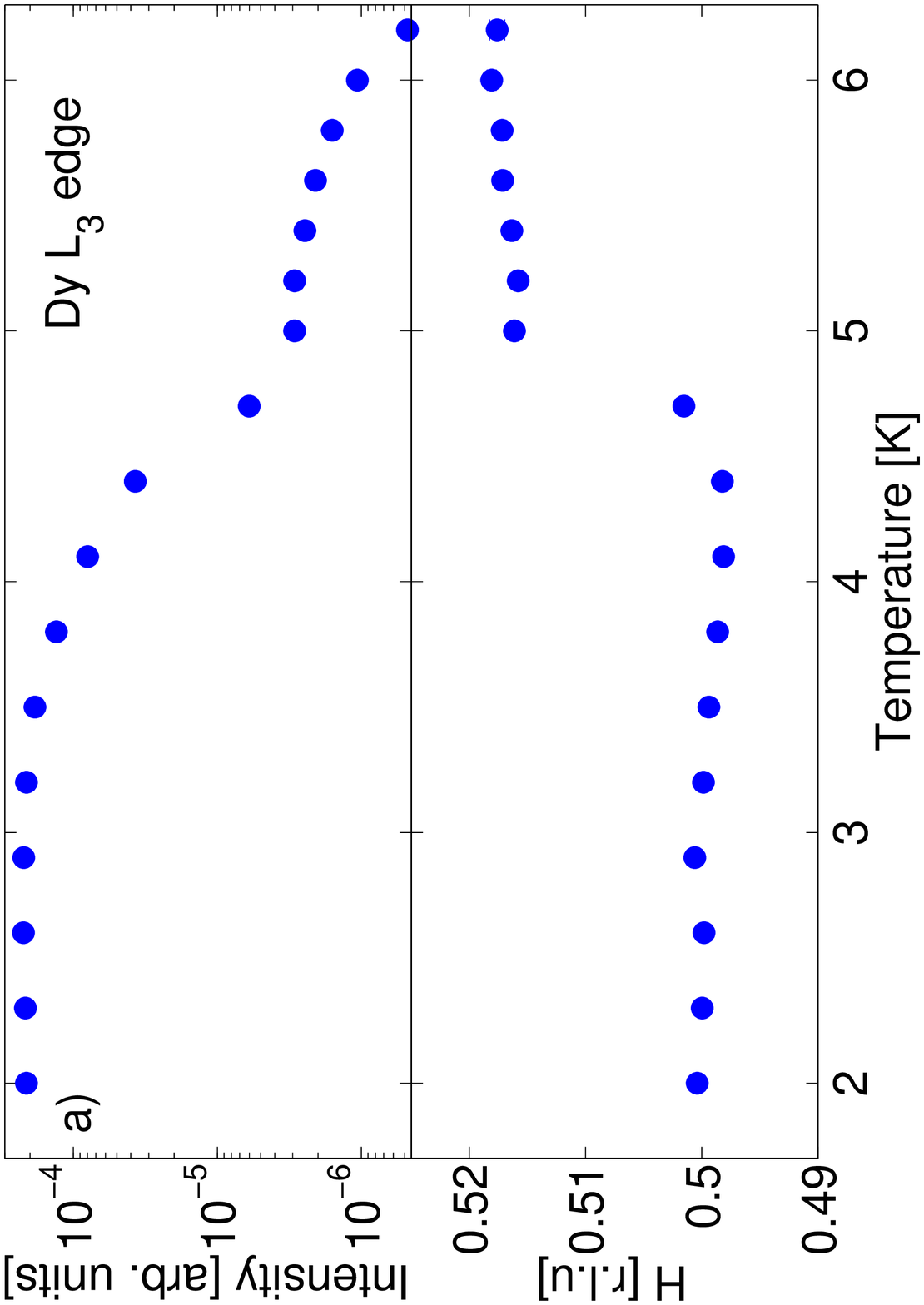}}
     %\hspace{.1in}
     \subfigure{
          \includegraphics*[scale=0.35]{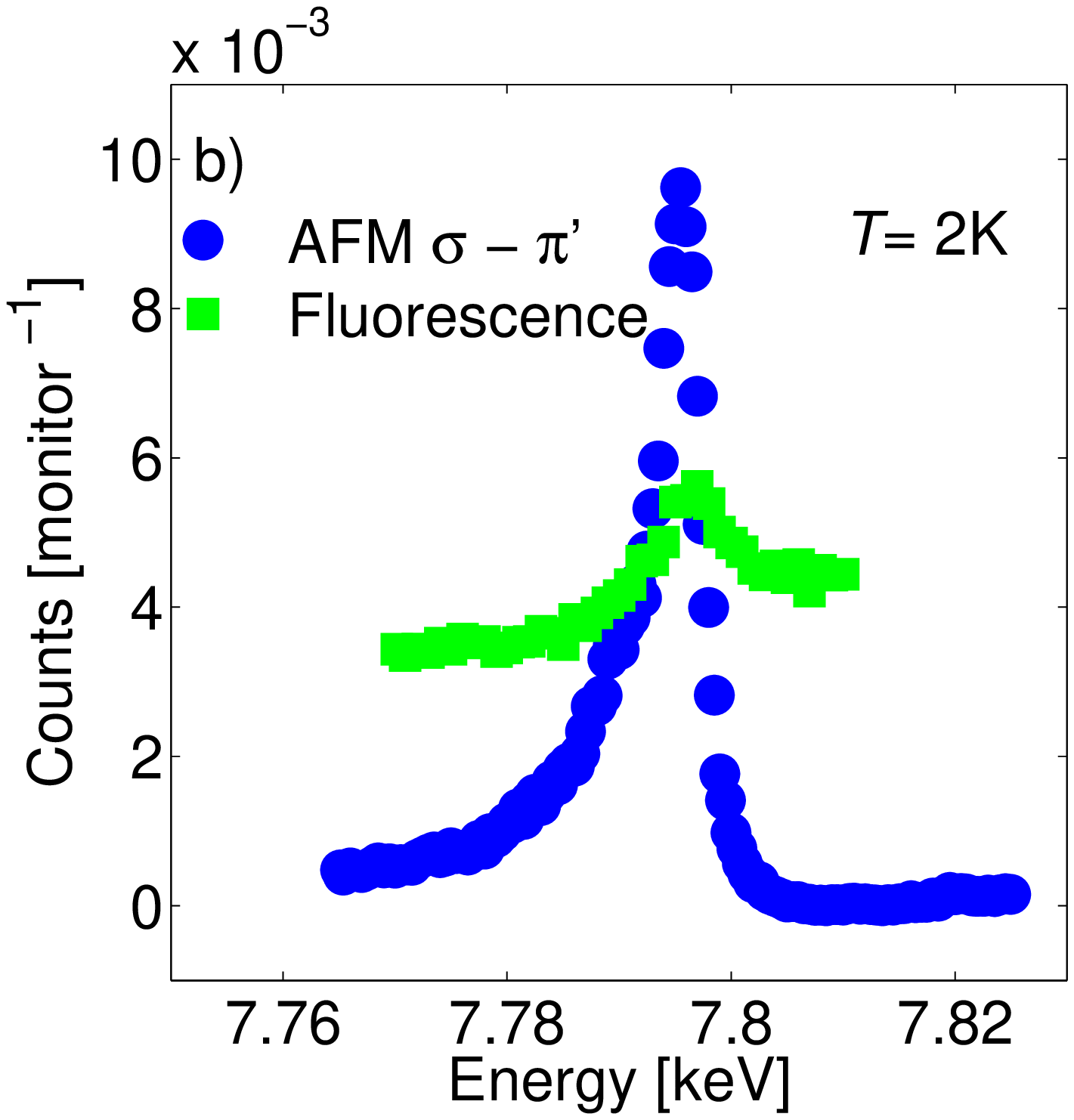}}\\
     \caption{(Color online) XRS from the AFM ordering of the Dy ions in DyMn$_{2}$O$_{5}$ with wavevector $\mathbf{q}_{AFM}^{Dy}=(0.5,0,0)$.
     (a) Results of fits to the peak in scans parallel to $(H,0,0)$
        to a Gaussian lineshape. Notice that the intensity is plotted on a logarithmic scale, and that the incommensurate
        signal is two orders of magnitude weaker than the AFM signal. Above 6.2K the
        intensity became so weak as to be below the detection
        threshold of the apparatus.
        (b) Scans of the incident x-ray energy at fixed wavevector $\mathbf{q}_{AFM}^{Dy}=(0.5,0,0)$, showing a strong
        resonance at the Dy L$_{3}$ edge (blue
        triangles) and the sample fluorescence (green
        squares, rescaled).}\label{Dy AFM and discomm Int,H}
\end{figure}

The first XRS measurements we describe were designed to confirm the
existence of antiferromagnetic (AFM) order on the Dy ions at low
temperatures, an effect which is seemingly ubiquitous in the {\it
R}Mn$_{2}$O$_{5}$ compounds when {\it R} is magnetic. Figure \ref{Dy
AFM and discomm Int,H}(a) shows the integrated intensity and
position in reciprocal space of the scattering arising from the Dy
AFM order. On warming, the intensity of the scattering decreases
monotonically, as expected, until $T_{\rm N}^{Dy}$ is reached.
However, between roughly 4.5\,K and 5\,K the ordering wavevector of
the Dy ions suddenly changes from $(0.5,0,0)$ to $(0.52,0,0)$. Such
an effect has not previously been observed in neutron or x-ray
scattering measurements on any of the {\it R}Mn$_{2}$O$_{5}$
compounds. Figure \ref{Dy AFM and discomm Int,H}(b) shows a scan of
the incident x-ray energy at the $(0.5,0,0)$ position at 2\,K. There
is a clear resonance at 7.795\,keV, which corresponds to virtual
transitions to the 5$d$ states. The resonance peak is broadened on
the left-hand side by about 10eV, and the background level on the
left-hand side is higher than that on the right-hand side. Such
broadening is probably due to interference between resonant and
non-resonant magnetic scattering \cite{Hill nonres res
interference,Neubeck nonres res interference}. Figure \ref{Dy AFM
and discomm Int,H}(b) also shows the sample fluorescence, which has
been rescaled because a different detector was used to measure it.

\begin{figure}[t]
\includegraphics*[scale=0.34,angle=270]{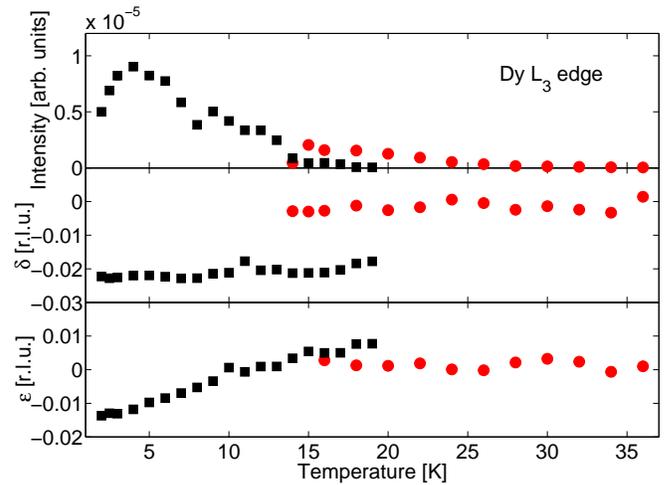}
\centering \caption{(Color online) XRS data from DyMn$_{2}$O$_{5}$
at the wavevectors $\mathbf{q}_{ICM}=(-0.5+\delta,0,0.25+\epsilon)$
(filled squares) and $\mathbf{q}_{CM}=(-0.5,0,0.25)$ (filled
circles). Diffraction peaks were fitted to a Gaussian lineshape for
scans parallel to $(H,0,0)$, and to a Lorentzian-squared for scans
parallel to $(0,0,L)$. The integrated intensity shown in the top
panel is the product of the Gaussian amplitude and width at each
temperature. The error bars are smaller than the size of the
points.}\label{3-pane IC and C Int,H,L}
\end{figure}

Figure \ref{3-pane IC and C Int,H,L} shows the temperature
dependence of the intensity and wavevector of the magnetic signal
associated with the order on the Mn sublattice. The scattering was
measured at several ICM wavevectors of the form $\mathbf{Q}=(0, 0,
4) + \mathbf{q}_{ICM}$ where $\mathbf{q}_{ICM}=(\pm0.5\pm\delta,
0, \pm0.25\pm\epsilon)$. At low temperatures this ICM phase has
non-zero $\delta$ and $\epsilon$. Between $T=2$\,K and $T=19$\,K
$\delta$ increases slightly from about $-0.023$ to $-0.018$, and
$\epsilon$ increases almost linearly from $-0.015$ to zero, then
changes sign and increases further to 0.007. The intensity of the
scattering has a maximum at 5\,K, the same temperature at which
the Dy AFM order disappears. Above 5\,K the intensity decreases
steadily, and eventually falls below the detection threshold at
$T=19$\,K.

Figure \ref{3-pane IC and C Int,H,L} also shows that for
$T\geq14$\,K a CM phase appears, i.e. one for which
$\mathbf{q}_{CM}=(0.5,0,0.25)$. The intensity of this signal grows
upon warming, becoming stronger than the ICM signal by 15\,K,
where it also reaches a maximum. The maximum intensity of the CM
signal is a factor of 5 weaker than the maximum intensity of the
ICM signal. As temperature is increased above 15\,K the intensity
of the scattering gradually decreases, falling below the detection
threshold for $T>37$\,K.

Signals were measured at the equivalent positions
$\mathbf{q}_{1}=(-0.5+\delta, 0, 0.25-\epsilon)$,
$\mathbf{q}_{2}=(-0.5-\delta, 0, 0.25-\epsilon)$ and
$\mathbf{q}_{3}=(-0.5+\delta, 0, -0.25+\epsilon)$ for several
temperatures. Although the same behaviour was observed at each
position the absolute intensity of the signal was strongest at
$\mathbf{q}_{2}$, probably due to a certain amount of
self-absorption of x-rays by the crystal in the other
orientations. The intensity and wavevector at each temperature
were determined by fitting the lineshape in scans parallel to
$(H,0,0)$ to a Gaussian, and scans parallel to $(0,0,L)$ to a
Lorentzian-squared function. These functions were chosen because
they gave the best fit to the data. The maximum intensities of
these signals were at least an order of magnitude smaller than the
signal arising from the AFM order of the Dy ions. To illustrate
this, Figure \ref{Logscale intensity IC,C,AFM} displays the
intensities of each type of magnetic peak as a function of
temperature, on a logarithmic scale.

\begin{figure}[!h]
\includegraphics*[scale=0.4,angle=0]{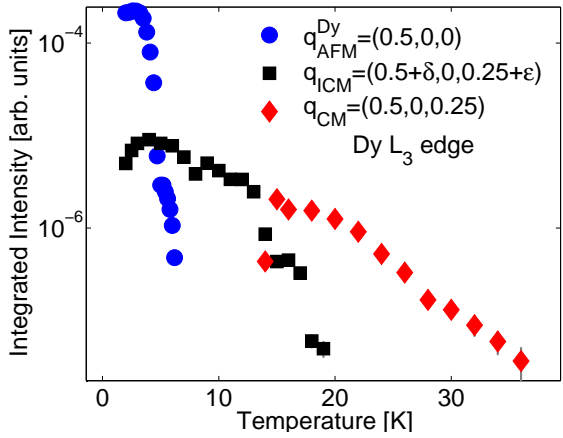}
\centering \caption{(Color online) Integrated intensity of the
Dy-AFM, ICM and CM signals, plotted on a logarithmic scale. The
coupling between Dy and Mn ions in the ICM and CM phases, given by
the order of magnitude of the intensity, appears to decreases
approximately as a power law with increasing
temperature.}\label{Logscale intensity IC,C,AFM}
\end{figure}

\begin{figure}[!h]
\includegraphics*[scale=0.4,angle=0]{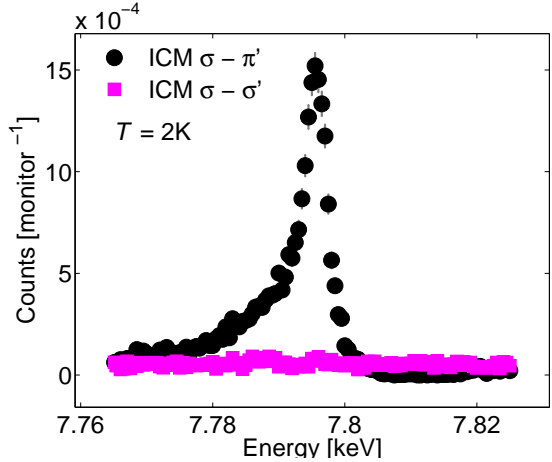}
\centering \caption{(Color online) Scans of the incident x-ray
energy for the ICM signal at $(0.5+\delta,0,0.25+\epsilon)$. The
strong peaks in the $\sigma - \pi'$ channel, together with a lack of
signal in the $\sigma - \sigma'$ channel, shows that the scattering
is purely magnetic at this wavevector.}\label{Energy scans}
\end{figure}

Figure \ref{Energy scans} shows scans of the incident x-ray energy
through the Dy L$_{3}$-edge at the ICM wavevector
$(0.5+\delta,0,0.25+\epsilon)$. The strong resonance shows that
the signal observed in this experiment is due to the magnetic
interaction `felt' by the electrons excited into the $5d$ states
caused by the magnetic order of the surrounding Mn ions. The
scattering is entirely in the $\sigma - \pi'$ polarisation
channel, with nothing in the $\sigma - \sigma'$ channel, which
shows that this ordering is magnetic and essentially dipolar,
since any quadrupolar term, which would appear in the $\sigma -
\sigma'$ channel, is immeasurably small. The broadening of the
resonance peak at energies below the Dy L$_{3}$ edge presumably
has the same origin as that discussed earlier in relation to
Figure \ref{Dy AFM and discomm Int,H}(b), i.e. interference
between non-resonant magnetic scattering and resonant magnetic
scattering.

\begin{figure}[!h]
\includegraphics*[scale=0.4,angle=0]{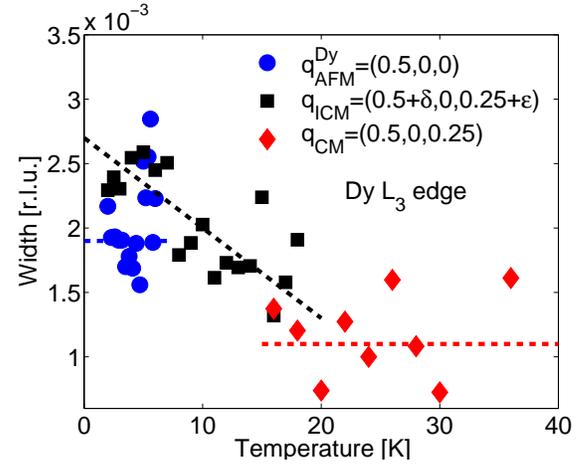}
\centering \caption{(Color online) The widths (FWHM) of the peaks
corresponding to the Dy AFM, ICM and CM order measured in scans
parallel to $(0,0,L)$. The widths were determined by moment
analysis. The dashed lines are guides to the eye. Similar analysis
of scans parallel to $(H,0,0)$ and $(0,K,0)$ (not shown) shows no
change in the width in these directions.}\label{L-widths C,IC,AFM}
\end{figure}

Figure \ref{L-widths C,IC,AFM} shows how the widths of the AFM,
ICM and CM peaks vary with temperature for scans parallel to
$(0,0,L)$. The peaks are fitted to a Lorentzian-squared lineshape.
It is clear that the correlation length, proportional to
$1/(\mbox{width})$, is shorter for the ICM order at low
temperatures than for the Dy AFM and the CM order at higher
temperatures. Indeed, the correlation length of the order on the
electrons spins in the $5d$ states of the Dy ions, induced by the
Mn sublattice ordering, increases with increasing temperature,
before becoming constant above about 20K.

\begin{figure}[!h]
\includegraphics*[scale=0.4,angle=0]{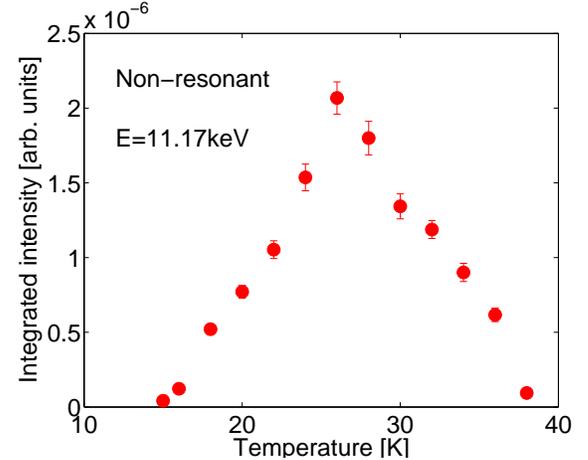}
\centering \caption{(Color online) Temperature variation of the
intensity of the signal at $\mathbf{q}=(0,0,0.5)$ measured with a
non-resonant x-ray energy. The intensities were calculated from fits
to a Gaussian for scans parallel to $(H,0,0)$. This scattering is
due to a change in the crystal structure.}\label{2Q Intensity}
\end{figure}

Figure \ref{2Q Intensity} shows the results of non-resonant x-ray
scattering measurements made with an incident x-ray energy of
11.17\,keV, plotting the intensity of the signal at
$\mathbf{Q}=(0, 0, 4.5) = (0, 0, 4) + 2\mathbf{q}_{CM}$, versus
temperature. At no temperature was a signal corresponding to
$(0,0,4) \pm 2\mathbf{q}_{ICM}$ observed. The signal at
$2\mathbf{q}_{CM}$ appears at $T=15$\,K and its intensity
increases steadily with warming, reaching a maximum at 27\,K. On
further warming the intensity decreases until it becomes too weak
to measure at $T>38$\,K. This measurement is in complete agreement
with the x-ray scattering data of Higashiyama {\it et al.}
\cite{Higashiyama} on the same material. Interestingly, the
temperature at which this signal is a maximum is the same
temperature at which there is a distinct anomaly in the specific
heat (see Figure \ref{My meas HC}), and at which the FE
polarisation reaches a local maximum \cite{Higashiyama}. Since this peak occurs in a %added ref
non-magnetic, non-resonant channel it must arise from Thompson
scattering, i.e. scattering of the x-rays by the charge of the
ions in the system, and therefore be structural in origin.

\section{Discussion}
XRS methods have been used to examine the magnetic ordering in
\DMO by measuring the magnetism on the Dy ions induced by the
magnetic environment of the surround Mn ions. Several observations
have been made which go beyond previous neutron scattering studies
of this compound. Specifically these are:

\begin{enumerate}

\item The existence of a structural distortion, characterised by
$\mathbf{q}=(0,0,0.5)$, for $15$\,K$ \leq T<40$\,K, coexistent
with the strongest FE polarisation.

\item The existence of magnetic interactions in the $5d$ states of
the Dy ions induced by the magnetism of the surrounding Mn
sublattice. This is an effect which is present for $T \leq T_{\rm
N} \approx 40$K, and not just below 8K ($T \leq T_{\rm N}^{Dy}$).

\item Changes in the $H$ component of the wavevector of the ICM
order for $2$\,K$ \leq T \leq 19$\,K, and indeed non-zero
incommensurability of this $H$ component, as well as an
incommensurate $L$ component.

\end{enumerate}

\noindent There have also been observed differences with XRS
measurements on the related compound TbMn$_{2}$O$_{5}$:

\begin{enumerate}

\item No sign of a re-entrant ICM phase at higher temperatures.

\item The existence of an ICM phase with negative $\epsilon$ as
well as positive $\epsilon$, as opposed to the phase with only
positive $\epsilon$ observed by Okamoto {\it et al.} \cite{Okamoto
TbMn2O5}.

\end{enumerate}

In addition, this study goes beyond all previous scattering
studies by measuring changes in the widths of the various peaks
with temperature. Finally, the existence of a previously
undetected magnetic ordering wavevector of $\mathbf{q}=(0.52,0,0)$
between 5\,K and $T_{\rm N}^{Dy}$ was observed.

There are, of course, many similarities between the findings in
this study and the findings of previous studies which used neutron
scattering to examine \DMO, or XRS to measure TbMn$_{2}$O$_{5}$:

\begin{enumerate}

\item The existence of two distinct phases below $T_{\rm N}$,
namely the ICM and CM phases.

\item Changes in the incommensurability of the ICM signal with
changing temperature.

\item The existence of antiferromagnetic order on the Dy ions at
low temperatures.

\end{enumerate}

%DISCOMMENSURATION OF THE DY-AFM:
The change in wavevector of the AFM order from $(0.5,0,0)$ to
$(0.52,0,0)$ shown in Figure \ref{Dy AFM and discomm Int,H}, which
has not been observed in neutron scattering measurements or indeed
in XRS measurements of other {\it R}Mn$_{2}$O$_{5}$ compounds,
might be explained as follows. The Dy $4f$ spins are coupled
antiferromagnetically, and so the exchange Hamiltonian contains
terms which are proportional to $\mu_{Dy}^{2}$, the square of the
ordered Dy magnetic moments. With increasing temperature the
ordered moment reduces, causing the exchange energy to decrease
rapidly. There is also an exchange interaction between the ordered
$4f$ spins and the $5d$ states, and this too weakens in the same
way. Coexistent with the Dy--Dy interaction there also exists a
coupling between the Dy $5d$ spin states and the Mn spins. The Mn
spins order at a much higher temperature and therefore their
average magnetic moment remains fairly constant at the low
temperatures under consideration. This means the exchange energy
of the Dy--Mn coupled system will reduce much less. At low
temperatures the Dy--Dy interaction dominates over the Dy--Mn
coupling, but as temperature increases the two couplings
eventually become similar in strength. This may allow the
$H$-component of the Dy $5d$ magnetic order to become entrained to
the ICM order of the Mn sublattice. Further support for such an
interpretation is found from the widths of the peaks in scans
parallel to $(0,0,L)$ shown in Figure \ref{L-widths C,IC,AFM}. The
change of wavevector from $(0.5,0,0)$ to $(0.52,0,0)$ is
accompanied by a sudden increase in the width to a value similar
to that of the ICM phase. If indeed the competition between Dy--Dy
AFM coupling and Dy--Mn ICM coupling causes the change in
wavevector then one might expect the width of the peak at
$(0.52,0,0)$ to be similar to that of the ICM peak at the same
temperature.

%WIDTH CHANGES:

To reiterate, the data presented show that the Dy $5d$ bands show
some magnetic order right up to the N\'{e}el temperature of the Mn
sublattice. Since the measurements presented in this paper show
only the magnetic order on the Dy $5d$ states there are two
possible ways to account for the changes in the width of the
L-component of the ICM and CM peaks, shown in figure \ref{L-widths
C,IC,AFM}. The first is that there is some magnetisation of the Dy
$4f$ electrons, which polarises the $5d$ states, and there is some
change in the correlation length of this. The second is that there
is a change in the magnetic order of the Mn which is then
reflected as a change in the Dy magnetic order. In the first case
it is possible that the correlation length of the magnetically
ordered Dy sublattice itself changes with temperature, with the Mn
correlation length remaining fixed, i.e. the Dy magnetic
correlation length does not reflect the Mn magnetic correlation
length. In the second case it is possible that the Dy $5d$ and Mn
magnetic correlation lengths behave in the same way as a function
of temperature, and it is the Mn correlation length that changes.
In either case the correlation length appears only to change along
the $c$-axis, and not along the $a$- or $b$-axes, evidenced by the
change in the width of the L-components but not in the H- or
K-components of the peaks (not shown).

One possible explanation of the increased $L$ width in the ICM phase
compared to the CM phase is that the magnetic structure is broken up
into locally commensurate domains between which there exist `slips'
which have the overall effect of making the magnetic structure
incommensurate. Koo {\it et al.} \cite{Koo TbMn2O5} suggest that in
TbMn$_{2}$O$_{5}$ the ICM phase may be interpreted as CM spin
modulations with domain walls, analogous to an effect observed in
ErNi$_{2}$B$_{2}$C \cite{Kawano-Furukawa ErNi2B2C}. The measurements
detailed above would appear to support such an interpretation, with
domain walls in the $ab$-plane.

%2Q SCATTERING
The non-resonant scattering measured at
$2\mathbf{q}_{CM}=(0,0,0.5)$, which is structural in origin, ties
in with a model proposed to explain multiferroicity in
TbMn$_{2}$O$_{5}$ \cite{Chapon TbMn2O5} and later extended to
describe YMn$_{2}$O$_{5}$ \cite{Chapon YMn2O5}. The {\it
R}Mn$_{2}$O$_{5}$ materials are geometrically frustrated, with
five different exchange interactions identified between the Mn
ions, and it is clear that one way to lift the resulting
degeneracy would be a small structural distortion. Analysis of the
atomic displacement parameters \cite{Chapon TbMn2O5} suggests that
a canted antiferroelectric (CAF) structural phase may be the way
in which this degeneracy is lifted. In this case the lattice
distorts in the opposite sense in adjacent unit cells along the
$c$-axis, but in the same sense along the $a$- and $b$-axes. In
each unit cell the distortions along the $a$-axis of different
ions are in different directions such that the FE polarisation
along the $a$-axis is cancelled out, however this does not occur
for the distortions along the $b$-axis. This would then give rise
to a FE polarisation along the $b$-axis, which is indeed what is
observed. A difficulty with this explanation of the occurrence of
ferroelectricity in DyMn$_{2}$O$_{5}$, however, is that the
amplitude of the scattering at $2\mathbf{q}_{CM}$ does not map on
to the amplitude of the observed FE polarisation. The polarisation
does have a local maximum at the same temperature at which the
scattering at $2\mathbf{q}_{CM}$ is most intense, $T_{max}=27$\,K,
but the onset of the largest polarisation is actually at about
17\,K, and the signal at $2\mathbf{q}_{CM}$ is relatively weak at
this temperature.

%COMPARISON WITH THEORY
Several phenomenological theories have been proposed to explain
the occurrence of multiferroicity \cite{Kenzelmann
PRL,Mostovoy,Betouras,Katsura theory}. A successful approach for
the {\it R}MnO$_{3}$ compounds has been developed by Mostovoy
\cite{Mostovoy} which uses symmetry considerations combined with a
direct magnetoelectric coupling. The theory shows that certain
types of spiral magnetic order can induce a FE polarisation. Such
an approach has proved very successful at explaining the
multiferroic behaviour of TbMnO$_{3}$, in particular. A modified
version of this approach has been suggested by Betouras {\it et
al.} \cite{Betouras} to account for the multiferroic properties of
the {\it R}Mn$_{2}$O$_{5}$-type compounds, in which the same
starting point of considering the free energy in a Landau model is
used, but with additional assumptions. These assumptions are that
the spin density wave (SDW) that couples to the FE polarisation is
\emph{acentric}, i.e. the spin density is not necessarily centred
on a lattice site and has a non-zero phase, and that the
polarisation and inverse ferroelectric susceptibility have small
oscillatory parts in addition to a constant term. The result of
this is that a spontaneous polarisation is only allowed for
magnetic phases which are \emph{commensurate}. It is clear that in
DyMn$_{2}$O$_{5}$ the polarisation is much larger in the CM phase,
so the present data tend to support this explanation.

As mentioned in the introduction, the magnitude of the FE
polarisation in DyMn$_{2}$O$_{5}$ is the strongest of all the
materials in the {\it R}Mn$_{2}$O$_{5}$ family. The results
presented in this paper show that there is a significant magnetic
interaction in the Dy $5d$ bands, and it is possible that there is
also an induced ordering of the partially occupied Dy $4f$ states.
It has been found previously that the size of the ordered moment
of the Dy ions in DyMn$_{2}$O$_{5}$ is larger than that of other
rare earth ions in the {\it R}Mn$_{2}$O$_{5}$ series \cite{Blake},
which might partly explain why such a clear signal at the Dy
L-edge resonance is observed here. The models used to explain
multiferroic effects discussed above do not constrain the
magneto-electric coupling to involve just the Mn ions, so in
principle magnetic order of the Dy ions with the same wavevector
as the magnetic order of the Mn ions could give rise to an `extra'
contribution to the FE polarisation through the same mechanism. If
such coupling were proportional to the size of the rare-earth
moment then that might explain the existence of the strongest
polarisation in DyMn$_{2}$O$_{5}$ and the weakest polarisation in
YMn$_{2}$O$_{5}$ among the {\it R}Mn$_{2}$O$_{5}$ materials.

\section{Conclusions}

In conclusion, the XRS measurements presented here have shown that
the magnetic ordering in DyMn$_{2}$O$_{5}$ bears many similarities
to that of other members of the {\it R}Mn$_{2}$O$_{5}$ family of
materials previously studied. These measurements also show extra
features not previously observed in neutron scattering measurements
on DyMn$_{2}$O$_{5}$. Of particular note is the observation that the
Dy $5d$ bands are magnetically polarised right up to $T_{\rm N} \sim
40$\,K, the N\'{e}el temperature of the Mn ions. If the $4f$ states
are magnetised in a similar fashion there may be an enhancement of
the ferroelectric polarisation in this material, given the large Dy
magnetic moment. Such measurements demonstrate that x-ray resonant
scattering is a powerful tool for studying materials in which there
exists more than one magnetic ion, in particular allowing one to
resonantly enhance the scattering from magnetic order on one
sublattice which has been induced by the magnetisation of the other
sublattice.

\section{Acknowledgements}

We thank F.R. Wondre for x-ray Laue alignment of the sample, P.J.
Baker for assistance with the heat capacity measurements, and S.J.
Blundell for useful discussions. We are grateful for financial
support from the Engineering and Physical Sciences Research
Council of Great Britain and from the Wolfson Royal Society
Research Merit Award scheme (DFM).

\end{document}